\documentclass[preprint,review,12pt]{elsarticle}
\usepackage[colorlinks=true,
            citecolor=blue,
            linkcolor=blue,
            urlcolor=blue]{hyperref}

\usepackage{amssymb}
\usepackage{amsmath}
\journal{Physica Medica}

\usepackage[
    a4paper,
    top=1.5cm,
    bottom=2cm,
    left=2.5cm,
    right=2.5cm
]{geometry}

\begin{document}

\begin{frontmatter}

%% Title
\title{Estimation of Very High Energy Electron (VHEE) beam parameters using an in-air screen}

%% Author list
\author[a]{Joel Valerian}
\ead{joel.valerian@student.unimelb.edu.au}
\author[a,b]{Paul Giansiracusa}
\author[a]{Vincent Lu}
\author[a]{Paarangat Pushkarna}
\author[a]{Roger Rassool}
\author[a,c]{Suzie Lyn Sheehy}
\author[b]{Yaw-Ren Eugene Tan}
\author[a]{Matteo Volpi}

\affiliation[a]{
organization={The University of Melbourne},
city={Melbourne},
country={Australia}}
\affiliation[b]{
organization={Australian Synchrotron - ANSTO},
city={Melbourne},
country={Australia}}
\affiliation[c]{
organization={ANSTO},
city={Sydney},
country={Australia}}

%% Abstract
\begin{abstract}

\textbf{Purpose:} Very high energy electron (VHEE) beams offer several promising properties for radiotherapy. Accurate beam delivery requires a beam model that accounts for scattering along the beam propagation path. This work develops and validates a beam model incorporating Multiple Coulomb Scattering (MCS) to estimate beam parameters.

\textbf{Methods:} MCS is incorporated as an additive term to the covariance matrix transformation for a drift space, allowing conventional quadrupole scans to estimate beam parameters in the presence of a downstream scatterer. The scattering correction factor is fitted to Monte Carlo (MC) simulations for a range of materials and thicknesses. The model is then validated by recovering the initial beam parameters from simulated quadrupole scans of a realistic beamline. Finally, the method is applied to in-air screen measurements at the Pulsed Energetic Electrons for Research (PEER) end station of the Australian Synchrotron.

\textbf{Results:} After fitting the correction factors, the mean and standard deviation of the estimation errors for the final emittance, diameter, divergence, and mean kinetic energy were $(-0.06 \pm 3.17)\%$, $(0.15 \pm 0.66)\%$, $(0.18 \pm 2.18)\%$, and $(-0.03 \pm 0.06)\%$, respectively. Simulated quadrupole scans successfully recovered the initial beam parameters, with errors of at most 13\%. PEER measurements yielded emittance estimates ranging from 0.238 to 1.753~mm.mrad across scans, which were within the expected range.

\textbf{Conclusion:} Following an initial quadrupole scan, the method allows the in-air beam size to be estimated for other quadrupole configurations. Further work is required to improve the consistency and reliability of the experimental estimates.

\end{abstract}

% %Graphical abstract

% \begin{highlights}
% \end{highlights}

% %% Keywords
% \begin{keyword}
% %% keywords here, in the form: keyword \sep keyword

% %% PACS codes here, in the form: \PACS code \sep code

% %% MSC codes here, in the form: \MSC code \sep code
% %% or \MSC[2008] code \sep code (2000 is the default)

% \end{keyword}

\end{frontmatter}

\clearpage
% \linenumbers

\section{Introduction}
\label{introduction}
Very high energy electron (VHEE) therapy is an emerging radiotherapy modality that uses electron beams with energies of approximately 100--250~MeV. The idea was first suggested by \citet{desrosiers_vhee_2000}, based on Monte Carlo (MC) simulations showing penetration depths comparable to X-rays. Initially, VHEE beams attracted limited interest because they offered few advantages over X-rays and were more difficult to produce. Interest grew following the discovery of the FLASH effect by \citet{favaudon_flash_2014}, whereby irradiation at ultra-high dose rates (UHDR) can improve healthy tissue sparing. This effect has since been observed in experiments involving various animal models and organs \citep{vozenin_flash_2019}. However, implementing FLASH therapy with X-rays is challenging because the high-current electron beams used to generate X-rays via a conversion target lose much of their energy as heat rather than producing usable X-rays, necessitating specialised conversion targets \citep{esplen_design_2022}. In contrast, VHEE beams can more readily deliver UHDR irradiation \citep{ronga_vhee_2021}.

The charged nature of electrons allows VHEE beam trajectories to be manipulated using electromagnetic fields. Compared with clinical proton beams in the energy range of 70--250~MeV \citep{mohan_proton_2017}, VHEE beams are easier to deflect. For example, a 250~MeV electron beam has approximately 70\% of the magnetic rigidity of a 70~MeV proton beam. Magnetic rigidity quantifies a charged particle's resistance to deflection in a magnetic field, with lower rigidity corresponding to greater deflection for a given field. Several studies have investigated the potential benefits of focusing VHEE beams \citep{kokurewicz_focused_2019,whitmore_focused_2021,fan_focused_2023,an_focused_2024,assalmi_focused_2025,valerian_focused_2025}, with preliminary experimental demonstrations reported by \citet{kokurewicz_experimental_2021} and \citet{whitmore_experimental_2024}.

One of the major barriers to VHEE therapy is the limited availability of suitable accelerator facilities for VHEE research. Efforts to address this limitation include the development of compact linear accelerators (linacs) based on C-band (6~GHz) and X-band (12~GHz) high-gradient radiofrequency (RF) technology, which are designed to reduce the accelerator footprint and facilitate installation in hospital bunkers \citep{fischer_vhee_2024,barty_design_2024}. CERN has also published a technical design report for an accelerator intended for FLASH-VHEE radiotherapy, providing a foundation for further developments in this area \citep{angoletta_deft_2025}.

In parallel with these accelerator development efforts, access to existing facilities for VHEE research is expanding. One such facility is the CERN Linear Electron Accelerator for Research (CLEAR) at CERN \citep{corsini_vhee_2024}, where a second beamline has recently been developed and equipped with large-aperture magnets for experiments involving focused VHEE beams \citep{dyks_design_2022}. Another relevant facility is the Pulsed Energetic Electrons for Research (PEER) end station at the Australian Synchrotron, which uses the synchrotron injector linac for VHEE research during machine study sessions \citep{cayley_design_2026}.

Accurate beam modelling is essential for estimating the spatial dose distribution delivered to a target. At its simplest, this requires characterising the transverse beam profile and propagating it to the target, such as a patient computed tomography (CT) scan or a water phantom. Existing VHEE studies commonly represent the beam using a Gaussian finite-source model defined downstream of the vacuum window, with subsequent transport modelled using either MC-based or kernel-based calculations \citep{sitarz_tps_2025}. Although practical, this approach has two limitations. First, defining the beam downstream of the vacuum window excludes scattering within the window as electrons transition from the linac vacuum into the air drift space. Second, the model does not account for changes in the upstream linac configuration that affect the beam delivered to the target. For example, adjusting an upstream quadrupole changes the beam size at the vacuum window and, consequently, the beam profile at the target. A model that accounts for both vacuum window scattering and the upstream beamline configuration is therefore needed to represent beam transport more accurately.

Transverse beam characterisation has been studied extensively at accelerator facilities, including synchrotrons, colliders, storage rings, and free-electron lasers. A common approach uses the covariance matrix formalism (CMF), which describes a beam using the second moments of transverse particle positions and momenta, thereby providing a statistical representation of its phase space \citep{bryant_transfer_1984}. These moments can be expressed in terms of Courant--Snyder (CS) parameters and measured using various techniques, including the widely used quadrupole scan method and its variations \citep{carlsten_rotated_1999,green_quadscan_2015,rossi_quadscan_2016,ma_quadcan_2019,kallestrup_dispersive_2021,roussel_auto_2023,dhital_auto_2024,kim_recon_2024,petersson_emittance_2025}. However, quadrupole scans require a suitable in-vacuum screen, which may not always be available.

This work extends the conventional quadrupole scan method by incorporating Multiple Coulomb Scattering (MCS) into the CMF model. Previous studies have investigated the effects of MCS on the beam covariance matrix and, consequently, on the CS parameters \citep{reid_growth_1991,maier_cmf_1998,muller_cmf_2001,durieu_strip_2001,mohl_source_2006}. MCS primarily increases beam divergence, leading to emittance growth. The present work focuses on transverse emittance, which characterises the area occupied by the beam in transverse phase space and indicates its ability to maintain a small transverse size during propagation.

The present approach is further motivated by practical limitations, particularly the lack of suitable in-vacuum diagnostics. For example, the PEER beamline has only one in-vacuum screen, located downstream of a dipole magnet, where dispersion complicates beam size measurements. In compact VHEE linacs, where space is limited, installing an in-vacuum screen can extend the beamline by at least 10~cm and increase maintenance requirements, as servicing the screen necessitates venting the linac vacuum. In addition, in-air diagnostics offer greater flexibility for modification and reconfiguration than in-vacuum systems. A method for estimating beam parameters from in-air measurements therefore provides an attractive and practical alternative.

First, an empirical correction factor is introduced into the characteristic scattering angle formula by fitting it to MC simulations performed using OpenTOPAS \citep{faddegon_topas_2020}. The corrected formula is then incorporated into the CMF model as an additive term in the covariance matrix transformation for a drift space, allowing the propagation of CS parameters through a homogeneous scatterer. By combining this modified CMF model with the quadrupole scan method, the initial CS parameters can be estimated from a quadrupole scan. The method is first evaluated using simulated quadrupole scans of a realistic beamline model based on PEER. Finally, the method is applied to in-air screen measurements acquired at the PEER end station to estimate the initial CS parameters.

\section{Methods}
\subsection{Scattering in the covariance matrix formalism}
\label{methods1}

The derivation is presented for the $x$-axis but applies equally to the $y$-axis. The coupling between the transverse axes is assumed to be negligible. The beam is characterised by the second moments of the transverse particle position $x$ and slope $x'$, represented by the $2\times2$ covariance matrix $\boldsymbol{C}$. The elements $C_{11}$ and $C_{22}$ are the position and slope variances, respectively, while $C_{12}$ is the position--slope covariance. The slope is defined as the ratio of transverse to longitudinal momentum and since the transverse momentum $p_x$ is generally much smaller than the longitudinal momentum $p_z$, the slope is also approximately the transverse angle $\theta$: 
\begin{equation}
    x' \equiv \frac{p_x}{p_z} = \tan(\theta) \approx \theta.
\end{equation}

The covariance matrix can be expressed in terms of the CS parameters $\beta$, $\alpha$, and $\gamma$, and the emittance $\varepsilon$:
\begin{equation}
    \boldsymbol{C}
    \equiv \begin{pmatrix}
        C_{11} & C_{12} \\
        C_{12} & C_{22}
    \end{pmatrix}
    = \begin{pmatrix}
        \langle x^2 \rangle & \langle xx' \rangle \\
        \langle xx' \rangle & \langle x'^2 \rangle
    \end{pmatrix}
    = \varepsilon
    \begin{pmatrix}
        \beta & -\alpha \\
        -\alpha & \gamma
    \end{pmatrix}.
\end{equation}
The emittance is defined as the square root of the determinant of the covariance matrix:
\begin{equation}
    \varepsilon
    \equiv \sqrt{\det(\boldsymbol{C})}
    = \sqrt{C_{11}C_{22}-C_{12}^{2}}
\end{equation}
The beam diameter $D$ is defined as twice the standard deviation of the position distribution:
\begin{equation}
    D \equiv 2\sqrt{C_{11}} = 2\sqrt{\beta\varepsilon}.
\end{equation}
Similarly, the beam divergence $\Theta$ is defined as twice the standard deviation of the angular distribution:
\begin{equation}
    \Theta \equiv 2\sqrt{C_{22}} = 2\sqrt{\gamma\varepsilon}.
\end{equation}
Throughout this work, the terms covariance matrix, CS parameters, and beam parameters are used interchangeably when referring to the transverse description of the beam. The terms beam size and beam diameter are also used interchangeably to describe its size.

A key advantage of the CMF model is that beam propagation through individual beamline elements can be described by transforming the covariance matrix using the corresponding transfer matrix $\boldsymbol{M}$:
\begin{equation}
    \boldsymbol{C}_{\mathrm{after}}
    = \boldsymbol{M}\boldsymbol{C}_{\mathrm{before}}\boldsymbol{M}^{\mathsf T},
\end{equation}
where $\boldsymbol{M}^{\mathsf T}$ denotes the transpose of the transfer matrix. The transfer matrices for drift spaces, focusing quadrupoles (QF), and defocusing quadrupoles (QD) are given by
\begin{equation}
    \boldsymbol{M}_{\mathrm{drift}} =
    \begin{pmatrix}
        1 & L \\
        0 & 1
    \end{pmatrix},
\end{equation}
\begin{equation}
    \boldsymbol{M}_{\mathrm{QF}} =
    \begin{pmatrix}
        \cos(kL) & \dfrac{\sin(kL)}{k} \\
        -k\sin(kL) & \cos(kL)
    \end{pmatrix},
\end{equation}
\begin{equation}
    \boldsymbol{M}_{\mathrm{QD}} =
    \begin{pmatrix}
        \cosh(kL) & \dfrac{\sinh(kL)}{k} \\
        k\sinh(kL) & \cosh(kL)
    \end{pmatrix}.
\end{equation}
Here, $L$ denotes the length of the drift space or the effective length of the quadrupole, and $k$ is the quadrupole focusing strength which is proportional to the quadrupole current.

As the beam traverses a scattering medium, MCS adds a contribution to the final covariance matrix. For a homogeneous scattering slab, the covariance matrix downstream of the slab is
\begin{equation}
    \boldsymbol{C}_{\mathrm{after}}
    =
    \boldsymbol{M}_{\mathrm{drift}}
    \boldsymbol{C}_{\mathrm{before}}
    \boldsymbol{M}_{\mathrm{drift}}^{\mathsf T}
    + \boldsymbol{\Delta}.
\end{equation}
Here, $\boldsymbol{M}_{\mathrm{drift}}$ describes propagation through the slab in the absence of scattering, while $\boldsymbol{\Delta}$ represents the additional covariance introduced by MCS:
\begin{equation}
    \boldsymbol{\Delta}
    =
    \begin{pmatrix}
        \Delta_{11} & \Delta_{12} \\
        \Delta_{12} & \Delta_{22}
    \end{pmatrix}.
\end{equation}
The $\boldsymbol{\Delta}$ matrix is obtained by integrating individual contributions of successive momentum-dependent small-angle scattering kicks throughout the slab, following \citet{negra_mcs_1971}. The derivation of the $\boldsymbol{\Delta}$ matrix, including the momentum-loss correction and an additional material-specific scattering correction factor $\lambda$, is provided in~\ref{app:derive_delta}.

\begin{table}[hb]
    \centering
    \begin{tabular}{ll}
        \hline
        \textbf{Parameter} & \textbf{Values} \\
        \hline
        Kinetic energy [MeV] & 100 and 200 \\
        $\varepsilon$ [mm.mrad] & 0.1 and 1 \\
        $\beta$ [m] & 2 and 10 \\
        $\alpha$ & $-5$ and $5$ \\
        Material & Air, water, titanium, and lead \\
        Thickness, $L/X_0$ & Air: $10^{-4}$--$10^{-2}$; others: $10^{-3}$--$10^{-1}$ \\
        Thickness sampling & Seven logarithmically spaced values per material \\
        \hline
    \end{tabular}
    \caption{Beam parameters and scatterer configurations used in simulation to determine the correction factors.}
    \label{tab:scattering_config}
\end{table}

To improve model accuracy, homogeneous scattering slabs of different materials and thicknesses are simulated with various beam parameters, as summarised in~\autoref{tab:scattering_config}. The simulation results are used to determine each material correction factor by minimising the difference between the calculated and simulated emittance. The accuracy of the model is then evaluated by comparing four quantities before and after applying the correction: emittance, diameter, divergence, and mean kinetic energy. The estimation error in percentage used throughout this work is defined as
\begin{equation}
    \% \mathrm{Error} = \frac{\mathrm{Estimate} - \mathrm{Actual}}{\mathrm{Actual}} \times 100.
\end{equation}

\begin{figure}[!htb]
    \centering
    \includegraphics[height=0.3\textheight]{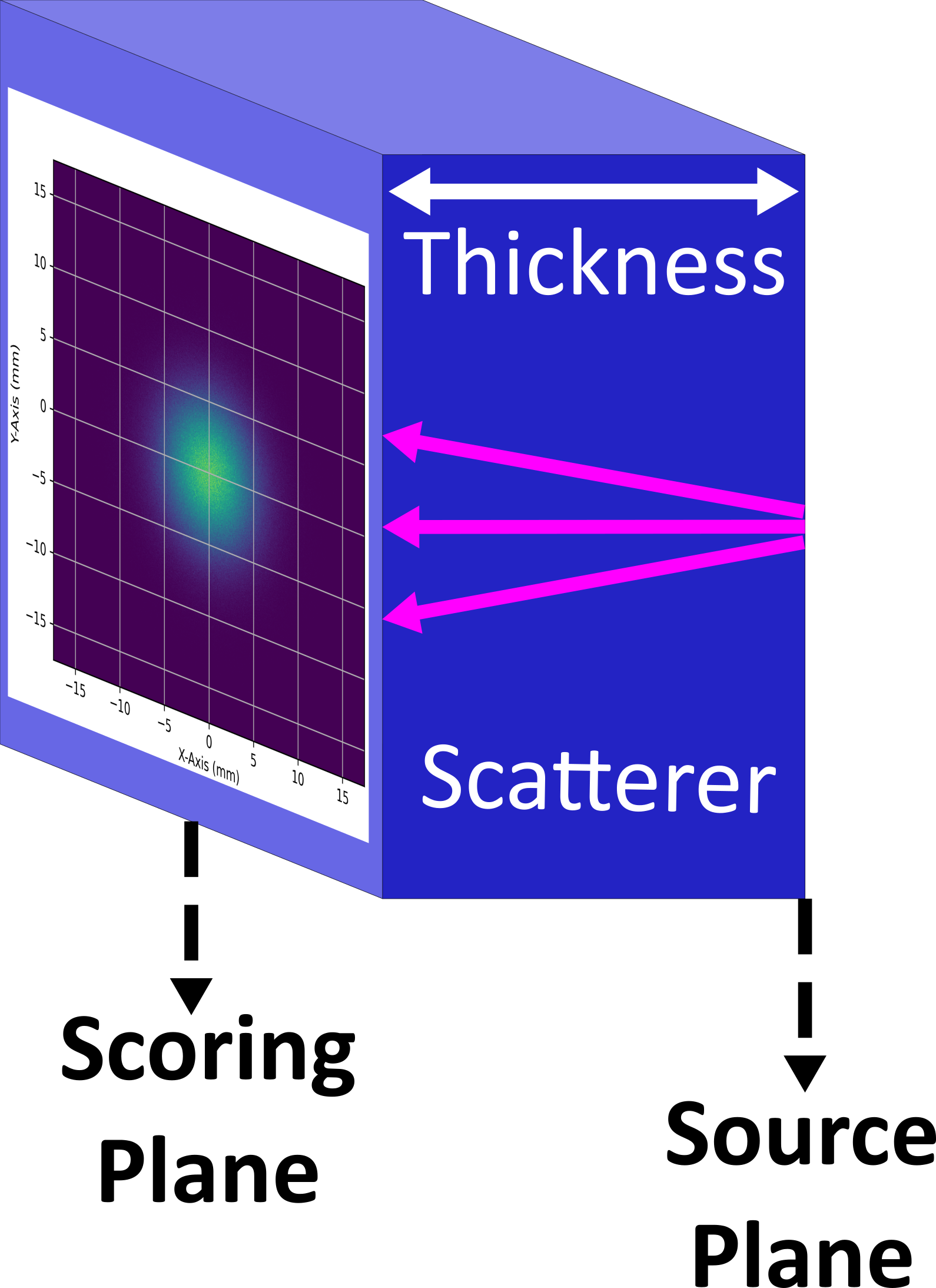}
    \caption{Simulation geometry (not to scale) used to determine the correction factors. The beam propagates from right to left, as indicated by the purple arrow. An example transverse beam profile is shown on the scoring plane at the left surface of the scatterer.}
    \label{fig:scattering_geometry}
\end{figure}

The simulation geometry is shown in~\autoref{fig:scattering_geometry}. An initial particle distribution is generated on the source plane from the specified initial kinetic energy and CS parameters, then transported through the scattering slab. The resulting phase space is scored to determine the final CS parameters.

\subsection{Simulated quadrupole scan}
\label{methods2}

\begin{figure}[!htb]
    \centering
    \includegraphics[width=\linewidth]{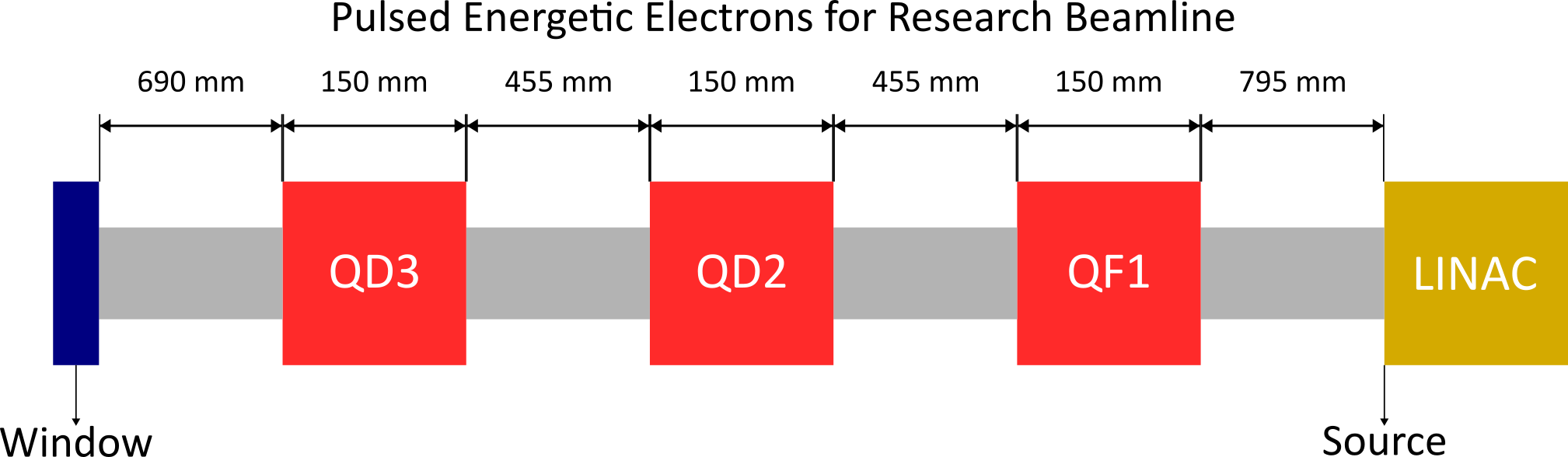}
    \caption{The diagram of the PEER beamline (not to scale) used in both the simulated quadrupole scans and CS parameters estimation at PEER. The in-air screen (not shown) is placed downstream of the window after 200~mm air drift for the simulation and 165~mm air drift for the PEER measurement. The window is made from 125~$\mu$m thick grade 2 Titanium. The beam travels from right to left.}
    \label{fig:peer_geometry}
\end{figure}

The realistic beamline model for the simulated quadrupole scans is based on the PEER beamline, which supplies VHEE beams to the PEER end station. The model begins at the linac exit and includes the three quadrupoles forming a triplet, a final vacuum drift, and a thin titanium vacuum window. The first quadrupole (QF1) focuses in the $x$-axis, whereas the second (QD2) and third (QD3) quadrupoles defocuses in the $x$-axis. The geometry of the beamline is shown in \autoref{fig:peer_geometry}. For the simulated quadrupole scan, a phase space scorer, which acts as the in-air screen in TOPAS, is defined after 200~mm of air drift downstream from the vacuum window.

The conventional quadrupole scan method assumes a simplified beamline consisting of a single thin scanning quadrupole followed by a drift space. The covariance matrix is obtained by fitting a second-order polynomial to the measured beam size, as described in~\ref{app:conventional_quadscan}. This approach is not directly applicable to the realistic beamline considered here, which comprises three thick quadrupoles, drift spaces of varying lengths, a titanium vacuum window, and an air drift space.

Instead, the initial CS parameters are estimated by minimising the difference between the measured beam diameter and those calculated using the CMF model of the full beamline as a function of the scanning quadrupole current. This approach accounts for the combined effects of the beamline elements and scattering on the final beam size. It also allows different quadrupoles to be scanned, rather than restricting the scan to the quadrupole immediately upstream of the screen, providing greater flexibility in scan configuration. 

An initial set of CS parameters is specified at the linac exit, located 795~mm upstream of QF1. The $\beta$ and $\alpha$ parameters in both transverse axes, together with the emittance, are selected arbitrarily, as listed in \autoref{tab:quadscan_parameters}. If the CMF with MCS model is valid, these initial CS parameters should be recoverable from the measured beam size curves at the screen.

\begin{table}[!htb]
    \centering
    \begin{tabular}{lcc}
        \hline
        \textbf{Parameter} & \textbf{Horizontal} & \textbf{Vertical} \\
        \hline
        $\beta$ [m] & 10.6 & 9.7\\
        $\alpha$ & 0.7 & -0.4 \\
        $\varepsilon$ [mm.mrad] & 0.8 & 0.8\\
        Kinetic energy [MeV] & \multicolumn{2}{c}{100} \\
        \hline
    \end{tabular}
    \caption{Beam parameters used in the simulated quadrupole scans. The CS parameters were chosen arbitrarily and does not represent the CS parameters at PEER.}
    \label{tab:quadscan_parameters}
\end{table}

Three scan configurations were investigated: \texttt{Q2SCANX}, \texttt{Q2SCANY}, and \texttt{Q3SCANXY}. For \texttt{Q2SCANX} and \texttt{Q2SCANY}, QF1 was held at a fixed current while QD2 was scanned, with QD3 switched off. For \texttt{Q3SCANXY}, QF1 and QD2 were held at fixed currents while QD3 was scanned. The quadrupole excitation coefficient at PEER, which relates the quadrupole focusing strength to the quadrupole current, was set to 0.429~T~m$^{-1}$~A$^{-1}$ for all three quadrupoles. The quadrupole current settings are summarised in \autoref{tab:quadscan_config}.

\begin{table}[!htb]
    \centering
    \begin{tabular}{lccc}
        \hline
        \textbf{Scan} & \textbf{QF1 (A)} & \textbf{QD2 (A)} & \textbf{QD3 (A)} \\
        \hline
        \texttt{Q2SCANX}  & -5.11 & $-8.00$ to $-5.00$ & 0.00 \\
        \texttt{Q2SCANY}  &  4.52 & $5.00$ to $8.00$ & 0.00 \\
        \texttt{Q3SCANXY} & -4.33 & -8.25 & $4.00$ to $10.00$ \\
        \hline
    \end{tabular}
    \caption{Quadrupole current configurations used for the simulated scans. Each scan comprised 31 uniformly spaced current settings, with increments of 0.1~A for \texttt{Q2SCANX} and \texttt{Q2SCANY}, and 0.2~A for \texttt{Q3SCANXY}.}
    \label{tab:quadscan_config}
\end{table}

The fixed quadrupole currents and scan ranges were selected with the aim of placing the beam size minimum near the centre of each scan. \texttt{Q2SCANX} and \texttt{Q2SCANY} were designed to locate the minimum in the horizontal and vertical axes, respectively, whereas \texttt{Q3SCANXY} was designed to locate the minima in both axes simultaneously.

Three beam diameter curves are compared: (1) simulation results obtained by fitting a Gaussian function to the phase space data; (2) CMF model estimations using the fitted CS parameters; and (3) CMF model estimations using the true CS parameters. The root-mean-squared errors (RMSE) between the simulated and fitted curves indicate the performance of the optimisation, whereas RMSE between the simulated and true curves indicate systematic discrepancies arising from either the CMF with MCS model or the processing of the MC phase space data. Ideally, all three curves should agree closely. The optimisation is also performed using the CMF without MCS model, in which all drift spaces, including the vacuum window and air gaps, are treated as vacuum. This comparison will highlights the importance of incorporating MCS when performing quadrupole scans using an in-air screen.

\subsection{Simulation setup}
\label{methods3}

All simulations were performed using OpenTOPAS (version 4.0.0), which is built on Geant4 (version 11.1.3), to model electron transport and interactions within the simulation geometry. The physics modules were selected in accordance with the recommendations of \citet{arce_geant4_2025} and the complete configuration is listed in \autoref{tab:simulation_parameters}.

\begin{table}[!htb]
    \centering
    \begin{tabular}{ll}
        \hline
        \textbf{Parameter} & \textbf{Setting} \\
        \hline
        MC Software & OpenTOPAS 4.0.0 (Geant4 11.1.3) \\
        Initial energy spread & 0 (monochromatic beam) \\
        Primary beam charge & 0.2 pC ($\sim1.25\times10^{6}$ particles)\\
        Physics modules &
            \begin{tabular}[t]{@{}l@{}}
                \texttt{g4em-standard\_opt4}\\
                \texttt{g4h-phy\_QGSP\_BIC\_HP}\\
                \texttt{g4decay}\\
                \texttt{g4ion-binarycascade}\\
                \texttt{g4h-elastic\_HP}\\
                \texttt{g4stopping}\\
                \texttt{g4radioactivedecay}
            \end{tabular} \\
        \hline
    \end{tabular}
    \caption{Simulation parameters and physics settings used in OpenTOPAS.}
    \label{tab:simulation_parameters}
\end{table}

The simulations were run on a system equipped with an AMD EPYC 9474F processor (16 virtual CPUs) and 64~GB of RAM. OpenTOPAS supports multithreaded execution, allowing particle histories to be processed concurrently. All available threads were used, and each simulation run could be completed in under a minute.

The VHEE beams were initialised with zero energy spread to represent monochromatic beams. Stochastic interactions as electrons traversed the scattering materials broadened the transmitted energy distribution and reduced the mean kinetic energy. Normalised emittance is generally required when comparing beams with different mean energies. However, the energy loss was expected to be small for most simulation runs, so only geometric emittance was used. The incident beam charge was set to $0.2~\mathrm{pC}$, corresponding to approximately $1.25\times10^{6}$ primary electrons, providing enough statistics for the phase space data.

The transverse particle distribution at the source can be defined using CS parameters to generate a Gaussian distribution. However, OpenTOPAS does not directly provide the CS parameters after particle transport. Instead, a phase space scorer is used to record particle positions and momenta at a specified scoring surface, typically located at the downstream surface of a scatterer. 

A Gaussian-fitting approach is used to calculate the CS parameters from the phase space data. First, particles with scattering angles exceeding $45^\circ$ are excluded to prevent outliers from disproportionately increasing the calculated emittance. The remaining phase space data are binned into a two-dimensional histogram with 1000 bins along each axis. The histogram is then fitted with a bivariate Gaussian function, $S(\boldsymbol{x})$, with the covariance matrix included among the fitting parameters:
\begin{equation}
    S(\boldsymbol{x}) = A\exp\left[-\frac{1}{2}(\boldsymbol{x} - \boldsymbol{\mu})^{\mathsf T}\boldsymbol{\Sigma}^{-1}(\boldsymbol{x}-\boldsymbol{\mu})\right],
\end{equation}
where $\boldsymbol{x}$ is the coordinate vector, $\boldsymbol{\mu}$ is the mean vector (first moment), $\boldsymbol{\Sigma}$ is the covariance matrix (second central moment), and $A$ is the amplitude. The covariance matrix is fitted independently in each transverse axis, $(x,x')$ and $(y,y')$, and the CS parameters are subsequently calculated from the fitted matrices. This approach provides CS parameters that can be compared directly with those used in the CMF model.

Finally, the quadrupole scan optimisation is performed using a standard least-squares curve-fitting algorithm. Bounds are imposed on the fitted $\beta$, $\alpha$, and emittance to constrain the optimisation to physically plausible values and prevent invalid CS parameters, such as zero emittance or combinations that yield a non-positive-definite covariance matrix. The correction factor applied to the characteristic scattering-angle formula also helps reduce small discrepancies between the simulation results and the CMF model.

\subsection{Experiment Setup}
\label{methods4}

The injector linac at the Australian Synchrotron is based on the designs of the Swiss Light Source and Diamond Light Source linacs \citep{schlott_linac_2000,christou_linac_2004}. It consists of several stages that generate, bunch, and accelerate the electron beam from 90~keV to 100~MeV. The linac operates in either single-bunch mode, producing a bunch shorter than 1~ns, or multi-bunch mode, producing trains of 20--100 bunches over a period of up to 200~ns. The measurements reported here were performed in multi-bunch mode, using a train of 100 bunches spanning 100~ns and carrying a total charge of approximately 1~nC.

During normal synchrotron operation, the bending dipole immediately downstream of the third quadrupole is energised, deflecting the beam within the vacuum pipe into a branch leading to the Linac-to-Booster (LTB) transfer line. During machine study sessions for PEER beamtime, the dipole is switched off, allowing the beam to travel straight through the beam pipe towards the 125~$\mu$m-thick titanium foil vacuum window and then the PEER end station, where equipment and samples can be installed. At the end station, the screen is mounted on a motorised linear stage and positioned 165~mm downstream of the vacuum window.

The beam profile is measured using a Blue 800 Speed X-ray intensifying screen designed for X-ray imaging. The resulting scintillation light is imaged using an Allied Vision Mako G-158 camera equipped with a Sony IMX273 monochrome CMOS global-shutter sensor. The sensor has a pixel size of $3.45\times3.45~\mu\mathrm{m}^2$, a resolution of $1456\times1088$ pixels, and a 12-bit pixel depth. The camera gain is set to zero, and the exposure time is set to 1~ms. The camera is synchronised with the injector linac and controlled through the EPICS control system. 

Exposure times as short as 16~$\mu$s are possible without reducing the recorded light intensity, as the scintillation is expected to occur on a nanosecond timescale, comparable to the approximately 100~ns beam train duration. However, shorter exposures require more precise synchronisation with the beam pulse. If the exposure window does not overlap with the beam passage, for example, the beam profile may be missed entirely. An exposure time of 1~ms relaxes this synchronisation requirement while keeping the background light level relatively low.

\begin{figure}[!htb]
    \centering
    \includegraphics[width=\linewidth]{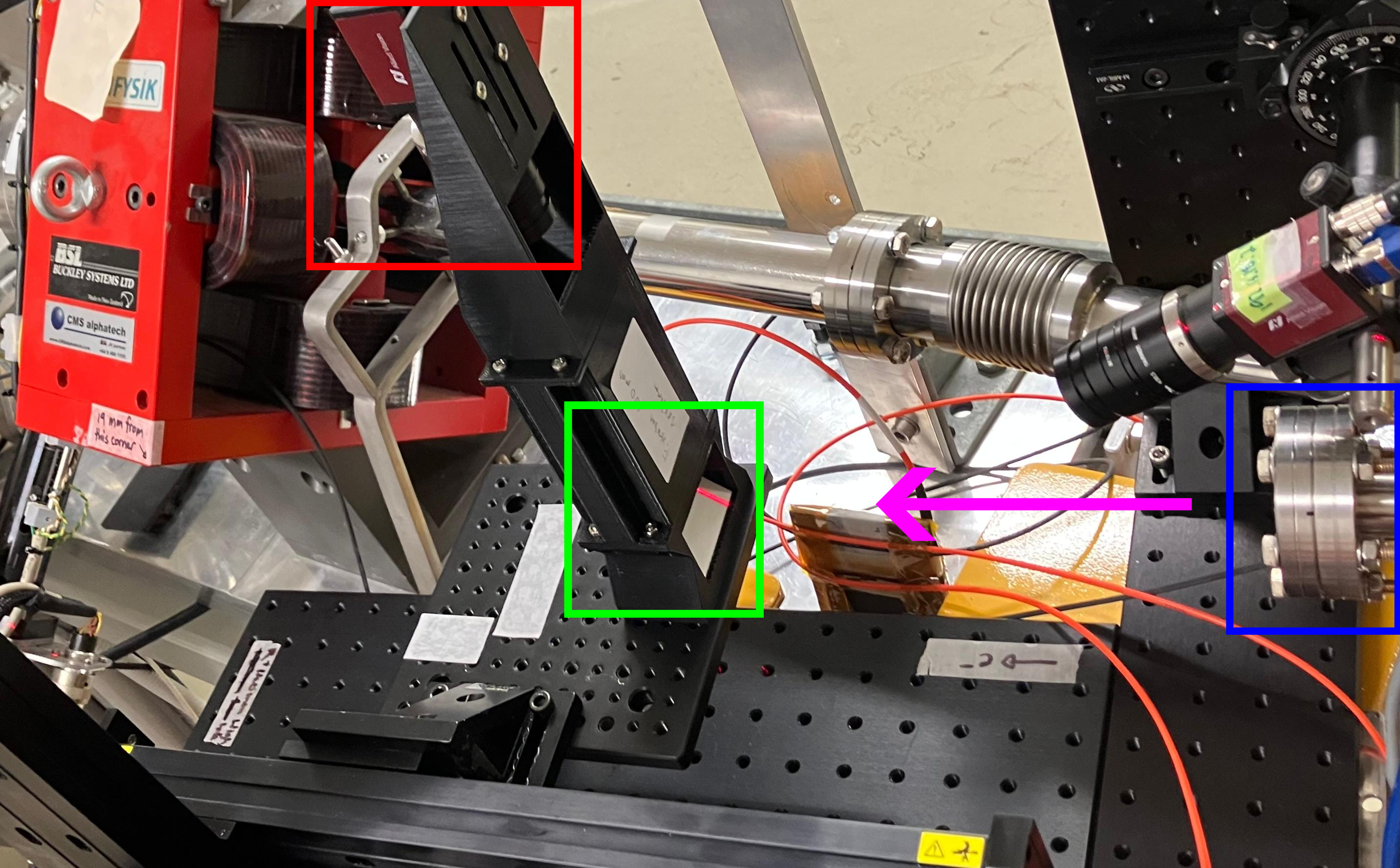}
    \caption{Experimental arrangement. The red box indicates the camera, the green box the scintillating screen, and the blue box the vacuum window. The purple arrow indicates the beam direction, from right to left. The custom holder (black), made of polylactic acid (PLA), secures the screen and camera in a fixed relative position.}
    \label{fig:experiment_setup}
\end{figure}

The screen is held in a custom 3D-printed holder that maintains the relative positioning of the camera and screen while reducing background light reaching the camera. The screen is mounted at an angle of $45^\circ$, facing both the incoming beam and the camera positioned above it. The arrangement is shown in \autoref{fig:experiment_setup}.

\begin{figure}[!htb]
    \centering
    \includegraphics[width=\linewidth]{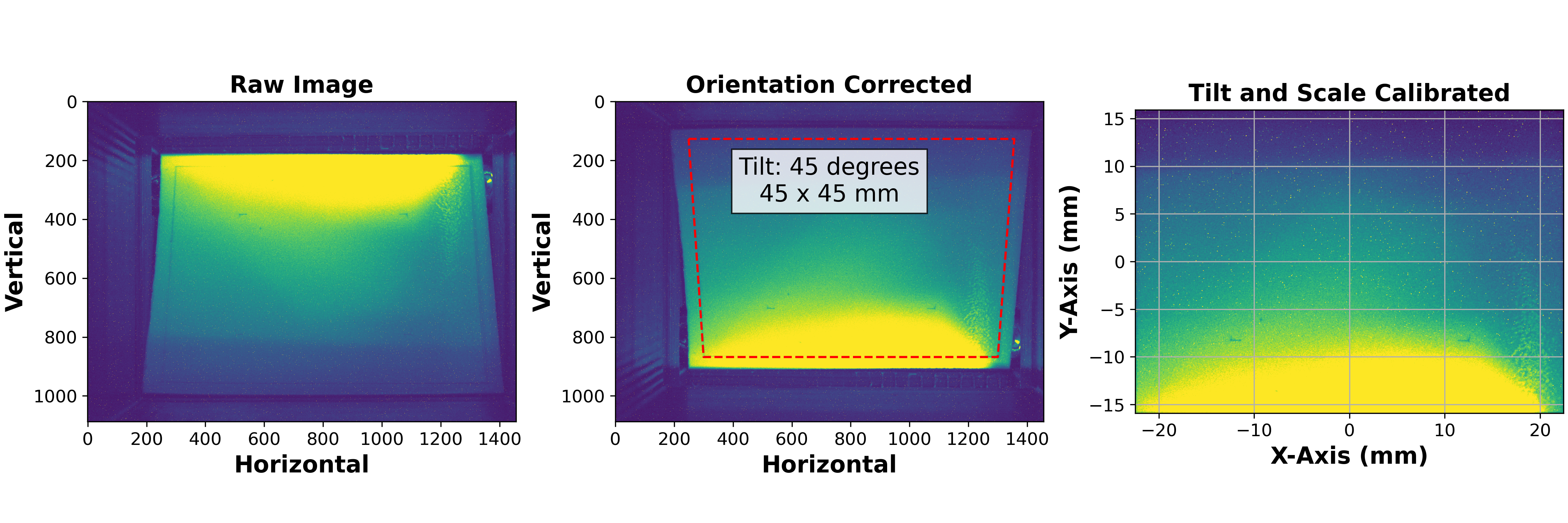}
    \caption{Screen image calibration and perspective correction. Left: original image. Centre: image after orientation correction. Right: image after perspective correction and scale calibration.}
    \label{fig:calibration}
\end{figure}

The beam profile image is first reoriented to align with the beam coordinate system, corresponding to a view looking down the beam pipe. A square of known dimensions ($45\times45~\mathrm{cm}^2$) is laser-engraved on the screen and used for calibration. A calibration image is acquired to determine the pixel scale and correct for perspective distortion and tilt (\autoref{fig:calibration}). A background image is then acquired and subtracted from each beam profile image. Finally, a Gaussian fit, consistent with the method used to calculate the CS parameters from the MC simulations, is applied to determine the beam size.

The quadrupole configurations used for the PEER measurements were as follows: (1) \texttt{MEASUREMENT\_X}, in which QF1 was scanned to minimise the beam size along the $x$-axis, with QD2 and QD3 switched off; (2) \texttt{MEASUREMENT\_Y}, in which QD3 was scanned to minimise the beam size along the $y$-axis, with QF1 and QD2 switched off; and (3) \texttt{MEASUREMENT\_XY}, in which QF1 was scanned to minimise the beam size along both axes simultaneously, with QD2 set to 9~A and QD3 switched off. Before each scan, the quadrupole currents were repeatedly cycled between 0~A and the maximum current of 14~A to minimise magnetic hysteresis. The current configurations are summarised in \autoref{tab:peer_config}. The CS parameters estimated from the PEER measurements are not expected to match the arbitrarily selected initial CS parameters used in the simulated quadrupole scans.

\begin{table}[!htb]
    \centering
    \begin{tabular}{lccc}
        \hline
        \textbf{Scan} & \textbf{QF1 (A)} & \textbf{QD2 (A)} & \textbf{QD3 (A)} \\
        \hline
        \texttt{MEASUREMENT\_X} & 3 to 7 & 0 & 0 \\
        \texttt{MEASUREMENT\_Y} & 0 & 0 & 5 to 9 \\
        \texttt{MEASUREMENT\_XY} & 4 to 12 & 9 & 0 \\
        \hline
    \end{tabular}
    \caption{Quadrupole current configurations used for the PEER measurements.}
    \label{tab:peer_config}
\end{table}

\section{Results}
\subsection{Scattering correction factor}
\label{results1}

\begin{figure}[!htb]
    \centering
    \includegraphics[width=\linewidth]{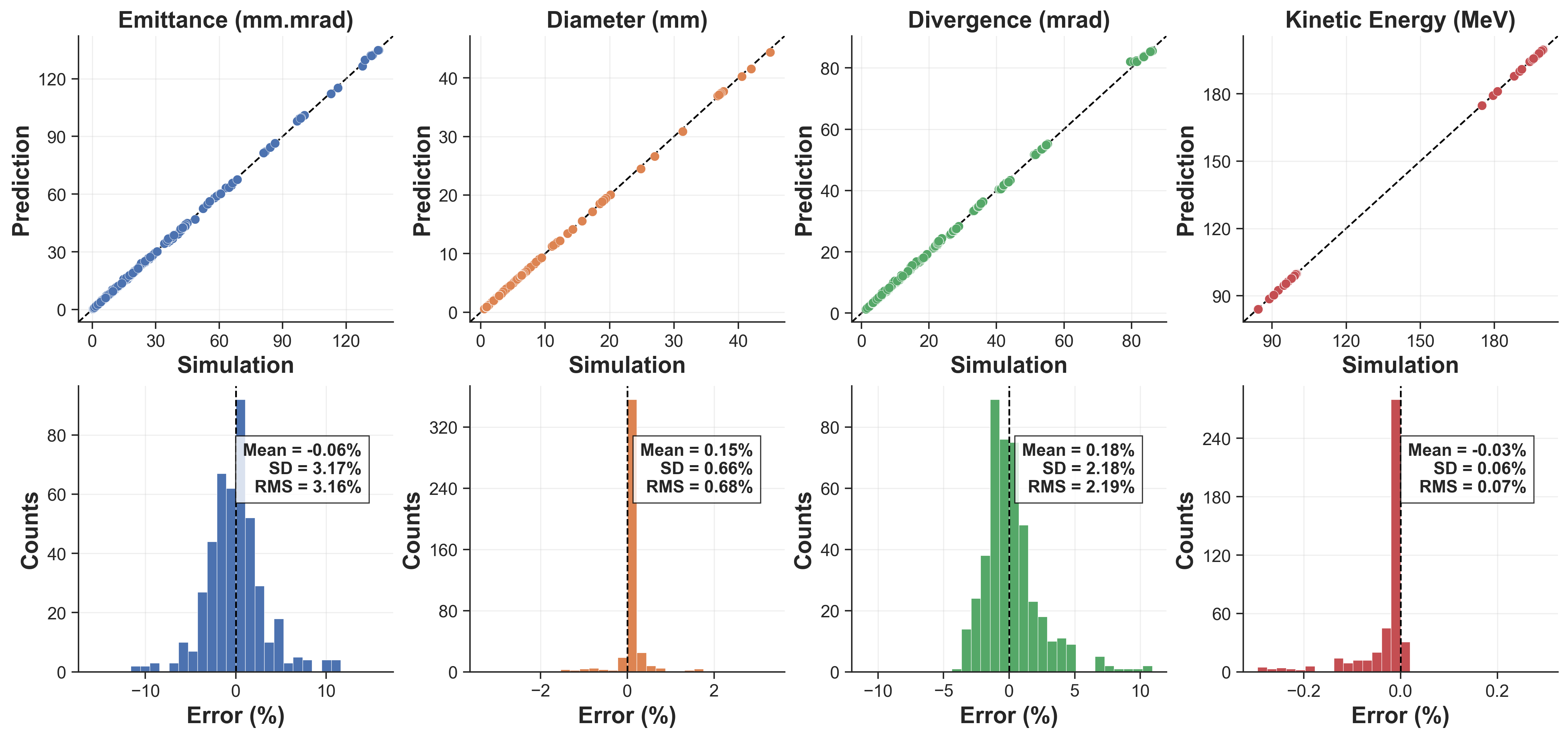}
    \caption{Top row: parity plots comparing CMF estimations with MC simulation results where points below the line are an underestimation while points above the line are an overestimation. Bottom row: distributions of the estimation errors in emittance, diameter, divergence, and mean kinetic energy.}
    \label{fig:scattering_result}
\end{figure}

The fitted correction factor for each simulated material is listed in \autoref{tab:correction_factor}. The mean and standard deviation of the estimation errors, before and after applying the correction factors, are summarised in \autoref{tab:scattering_result}. The parity plots and distributions of the estimation errors for emittance, diameter, divergence, and mean kinetic energy are shown in \autoref{fig:scattering_result}. The estimation errors are presented in percentage where negative and positive estimation errors indicates underestimation and overestimation from the CMF model, respectively.

\begin{table}[!htb]
    \centering
    \begin{tabular}{lc}
        \hline
        \textbf{Material} & \textbf{$\lambda$} \\
        \hline
        Air & 0.939  \\
        Water & 0.965 \\
        Titanium & 1.009 \\
        Lead & 1.039 \\
        \hline
    \end{tabular}
    \caption{The correction factor $\lambda$ for each material obtained from fitting the CMF with MCS model to MC simulations.}
    \label{tab:correction_factor}
\end{table}

\begin{table}[!htb]
    \centering
    \begin{tabular}{lcc}
        \hline
        \textbf{Parameter} & \textbf{Uncorrected error (\%)} & \textbf{Corrected error (\%)} \\
        \hline
        Emittance & $1.56\pm4.22$ & $-0.06\pm3.17$\\
        Diameter & $0.72\pm1.86$ & $0.15\pm0.66$ \\
        Divergence & $0.92\pm3.11$ & $0.18\pm
        2.18$ \\
        Mean kinetic energy & $-0.03\pm0.06$ & $-0.03\pm0.06$ \\
        \hline
    \end{tabular}
    \caption{The mean and standard deviation of the estimation errors across all simulation runs, in percentage, with the left and right column for before and after the correction factor, respectively.}
    \label{tab:scattering_result}
\end{table}

The fitted correction factors are close to unity where it is slightly below one for air and water, while those for titanium and lead are slightly above one, with at most 6\% difference for air. For all materials, applying the correction generally improves the estimations for all evaluated quantities except mean kinetic energy, which is essentially unchanged. The main quantity of interest is the emittance, in which the mean error percentage decreases from $(1.56\pm4.22)\%$ to $(-0.06\pm3.17)\%$.

Most data points in the parity plots lie close to the line of equality, indicating agreement between CMF with MCS and MC. In the rightmost parity plot, the two distinct clusters correspond to the 100 and 200~MeV beams. The emittance and divergence errors appear approximately normally distributed about their respective means, while the diameter errors are concentrated near zero. The mean kinetic energy error has a small negative tail, indicating occasional underestimation.

\subsection{Simulated quadrupole scan}
\label{results2}

\begin{figure}[!htb]
    \centering
    \includegraphics[width=\linewidth]{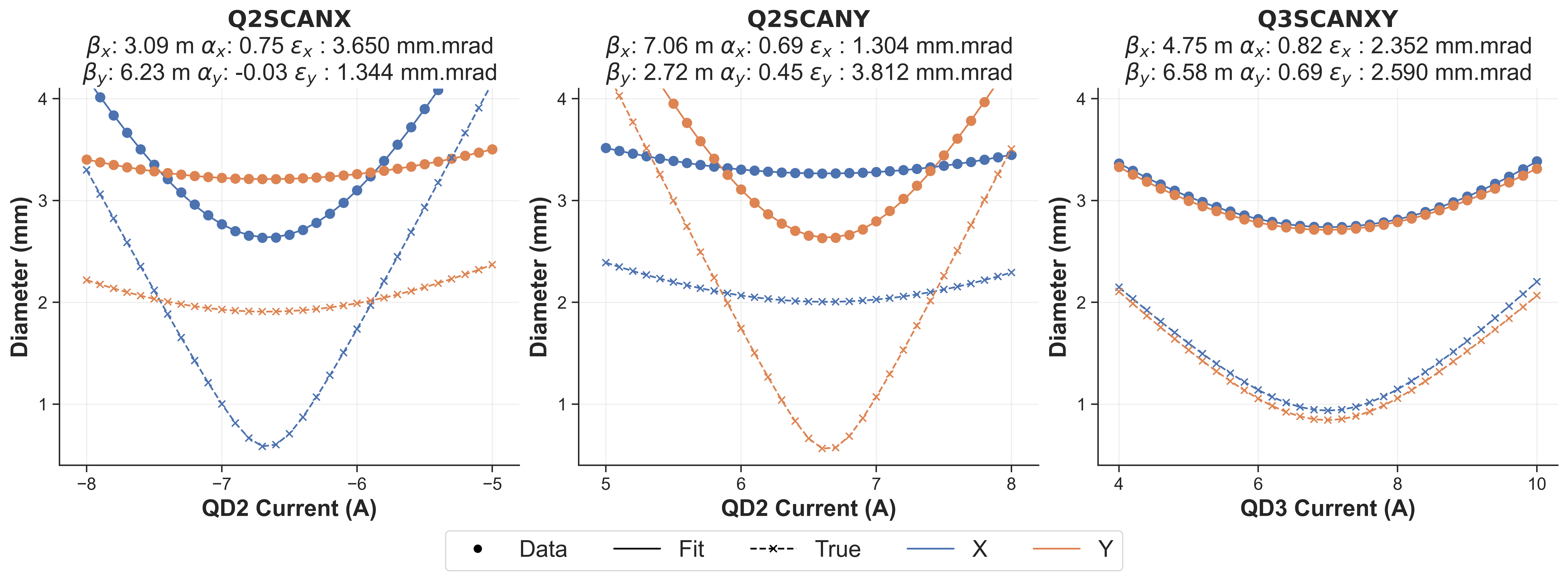}
    \caption{Simulated quadrupole scans using the CMF model without MCS effects. The fitted curves agree closely with the simulation data, whereas propagating the true CS parameters substantially underestimates the beam size.}
    \label{fig:quadscan_result_no_mcs}
\end{figure}

The simulated quadrupole scan results obtained using the CMF model without MCS effects are shown in \autoref{fig:quadscan_result_no_mcs}, and the optimisation results are summarised in \autoref{tab:quadscan_result_no_mcs}. The CMF model was still able to fit the simulated beam diameter curves, with data-to-fit RMSEs remaining below 6~$\mu$m across all scans, and estimate CS parameters. However, propagating the true CS parameters through the beamline without MCS effects resulted in substantial discrepancies between the estimated beam sizes and the MC simulation results. The CMF model consistently underestimated the beam sizes across the entire scan range, with data-to-true RMSEs reaching 1594~$\mu$m—approximately three orders of magnitude larger than the corresponding data-to-fit RMSEs across all axes and scans.

\begin{table}[!htb]
    \centering
    \begin{tabular}{lccc}
        \hline
        \textbf{Parameter} & \texttt{Q2SCANX} & \texttt{Q2SCANY} & \texttt{Q3SCANXY} \\
        \hline
        $\boldsymbol{\beta_x}$ [m]
            & $3.09$ ($-70.83\%$)
            & $7.06$ ($-33.43\%$)
            & $4.75$ ($-55.14\%$) \\
        $\boldsymbol{\alpha_x}$
            & $0.75$ ($7.51\%$)
            & $0.69$ ($-1.71\%$)
            & $0.82$ ($17.24\%$) \\
        $\boldsymbol{\varepsilon_x}$ [mm.mrad]
            & $3.650$ ($356.31\%$)
            & $1.304$ ($63.00\%$)
            & $2.352$ ($193.96\%$) \\
        \hline
        $\boldsymbol{\beta_y}$ [m]
            & $6.23$ ($-35.77\%$)
            & $2.72$ ($-71.99\%$)
            & $6.58$ ($-32.19\%$) \\
        $\boldsymbol{\alpha_y}$
            & $-0.03$ ($-92.01\%$)
            & $0.45$ ($-211.39\%$)
            & $0.69$ ($-271.92\%$) \\
        $\boldsymbol{\varepsilon_y}$ [mm.mrad]
            & $1.344$ ($67.94\%$)
            & $3.812$ ($376.55\%$)
            & $2.590$ ($223.69\%$) \\
        \hline
        \textbf{RMSE Fit, $x$} [$\mu$m]
            & $4.33$
            & $0.49$
            & $1.02$ \\
        \textbf{RMSE Fit, $y$} [$\mu$m]
            & $0.22$
            & $5.27$
            & $1.23$ \\
        \textbf{RMSE True, $x$} [$\mu$m]
            & $1399.96$
            & $1215.90$
            & $1541.64$ \\
        \textbf{RMSE True, $y$} [$\mu$m]
            & $1247.00$
            & $1386.38$
            & $1593.35$ \\
        \hline
        \hline
    \end{tabular}
    \caption{CS parameters estimated using the CMF model without MCS effects. Percentage errors relative to the true CS parameters are given in parentheses. The data-to-fit RMSE values are substantially smaller than the corresponding data-to-true RMSE values.}
    \label{tab:quadscan_result_no_mcs}
\end{table}

For \texttt{Q2SCANX}, the estimated emittance errors reach 357\% and 68\% in the $x$- and $y$-axes, respectively. Conversely, for \texttt{Q2SCANY}, the corresponding errors reach 63\% and 377\%. For \texttt{Q3SCANXY}, the emittance estimates are similarly inaccurate in both axes, with errors reaching 194\% in the $x$-axis and 224\% in the $y$-axis.

\begin{figure}[!htb]
    \centering
    \includegraphics[width=\linewidth]{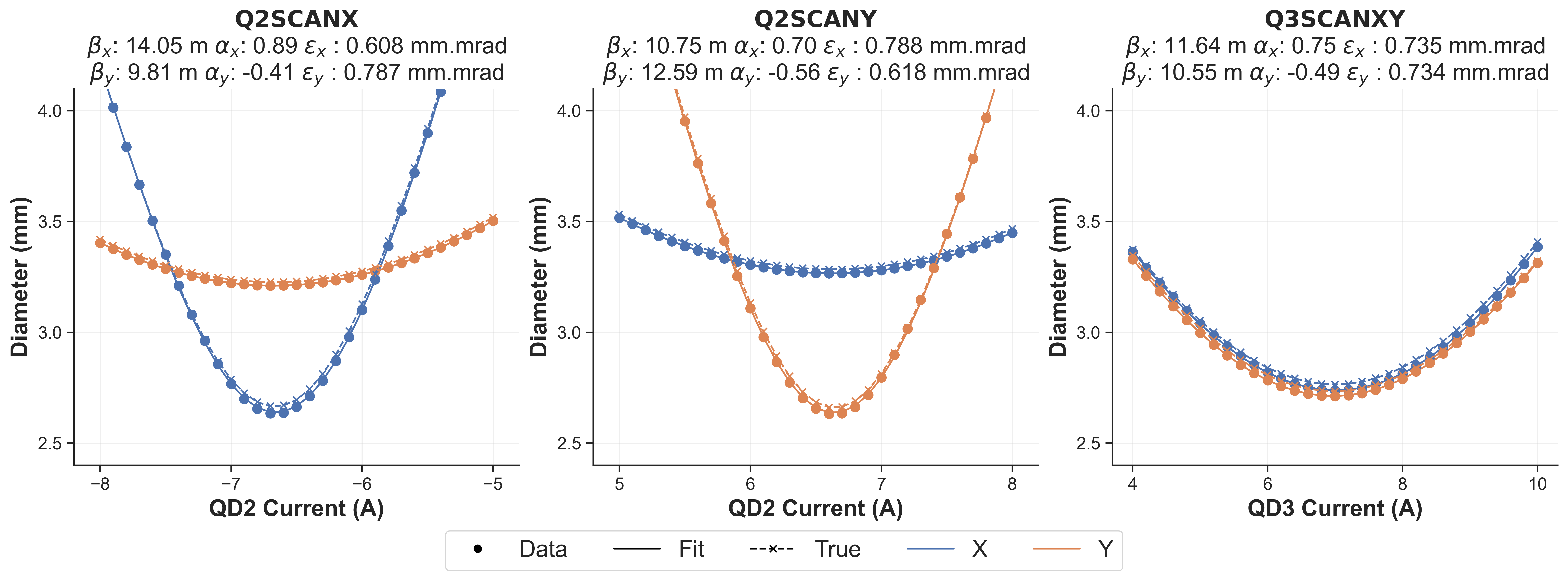}
    \caption{Simulated quadrupole scans using the CMF model with MCS effects. Both the fitted curves and those obtained using the true CS parameters agree closely with the simulation data, with slight deviations near the minima of the steeper curves.}
    \label{fig:quadscan_result_with_mcs}
\end{figure}

Incorporating MCS into the CMF model substantially improves the simulated quadrupole scan results. The beam diameter curves are shown in \autoref{fig:quadscan_result_with_mcs}, and the optimisation results are summarised in \autoref{tab:quadscan_result_with_mcs}. As in the case without MCS effects, the CMF model successfully fits the simulated beam-size curves, with data-to-fit RMSEs remaining below 6~$\mu$m across all scans. These RMSEs are essentially identical to those obtained without MCS effects. However, the data-to-true RMSEs are now below 23~$\mu$m, representing a reduction of approximately two orders of magnitude compared with the results obtained without MCS effects. This improvement is also evident in the beam diameter curves, where the fitted curves closely match those obtained using the true CS parameters.

\begin{table}[!htb]
    \centering
    \begin{tabular}{lccc}
        \hline
        \textbf{Parameter} & \texttt{Q2SCANX} & \texttt{Q2SCANY} & \texttt{Q3SCANXY} \\
        \hline
        $\boldsymbol{\beta_x}$ [m]
            & $14.05$ ($32.53\%$)
            & $10.75$ ($1.44\%$)
            & $11.64$ ($9.84\%$) \\
        $\boldsymbol{\alpha_x}$
            & $0.89$ ($27.01\%$)
            & $0.70$ ($0.54\%$)
            & $0.75$ ($6.59\%$) \\
        $\boldsymbol{\varepsilon_x}$ [mm.mrad]
            & $0.608$ ($-24.04\%$)
            & $0.788$ ($-1.46\%$)
            & $0.735$ ($-8.11\%$) \\
        \hline
        $\boldsymbol{\beta_y}$ [m]
            & $9.81$ ($1.14\%$)
            & $12.59$ ($29.76\%$)
            & $10.55$ ($8.72\%$) \\
        $\boldsymbol{\alpha_y}$
            & $-0.41$ ($2.52\%$)
            & $-0.56$ ($39.47\%$)
            & $-0.49$ ($22.50\%$) \\
        $\boldsymbol{\varepsilon_y}$ [mm.mrad]
            & $0.787$ ($-1.59\%$)
            & $0.618$ ($-22.78\%$)
            & $0.734$ ($-8.28\%$) \\
        \hline
        \textbf{RMSE Fit, $x$} [$\mu$m]
            & $4.33$
            & $0.49$
            & $1.03$ \\
        \textbf{RMSE Fit, $y$} [$\mu$m]
            & $0.22$
            & $5.27$
            & $1.23$ \\
        \textbf{RMSE True, $x$} [$\mu$m]
            & $21.42$
            & $17.81$
            & $22.85$ \\
        \textbf{RMSE True, $y$} [$\mu$m]
            & $15.90$
            & $19.60$
            & $18.68$ \\
        \hline
        \hline
    \end{tabular}
    \caption{CS parameters estimated using the CMF model with MCS effects. Percentage errors relative to the true CS parameters are given in parentheses. The \texttt{Q2SCANX} and \texttt{Q2SCANY} scans provide accurate estimates of the $y$- and $x$-axis parameters, respectively, whereas the remaining estimates exhibit moderate errors.}
    \label{tab:quadscan_result_with_mcs}
\end{table}

For \texttt{Q2SCANX}, the estimated $x$-axis emittance deviates from the true value by up to $-24\%$, whereas the $y$-axis emittance error remains within $-1.6\%$. Conversely, for \texttt{Q2SCANY}, the $x$-axis emittance error is as low as $-1.5\%$, while the $y$-axis error reaches $-23\%$. The \texttt{Q3SCANXY} scan provides more balanced estimates across both axes, with emittance errors of up to $-8.1\%$ and $-8.3\%$ in the $x$- and $y$-axes, respectively. Although some discrepancies remain, these results represent a substantial improvement over those obtained using the CMF model without MCS efffects.

All three scans, both with and without MCS effects, exhibit a beam size minimum, where the beam size initially decreases and then increases as the quadrupole current increases further. The fitted curves generally follow the MC simulation data, although visible deviations occur near the minima in the scans where the MCS effects is considered.

\subsection{In-air screen measurements}
\label{results3}

\begin{figure}[!htb]
    \centering
    \includegraphics[width=\linewidth]{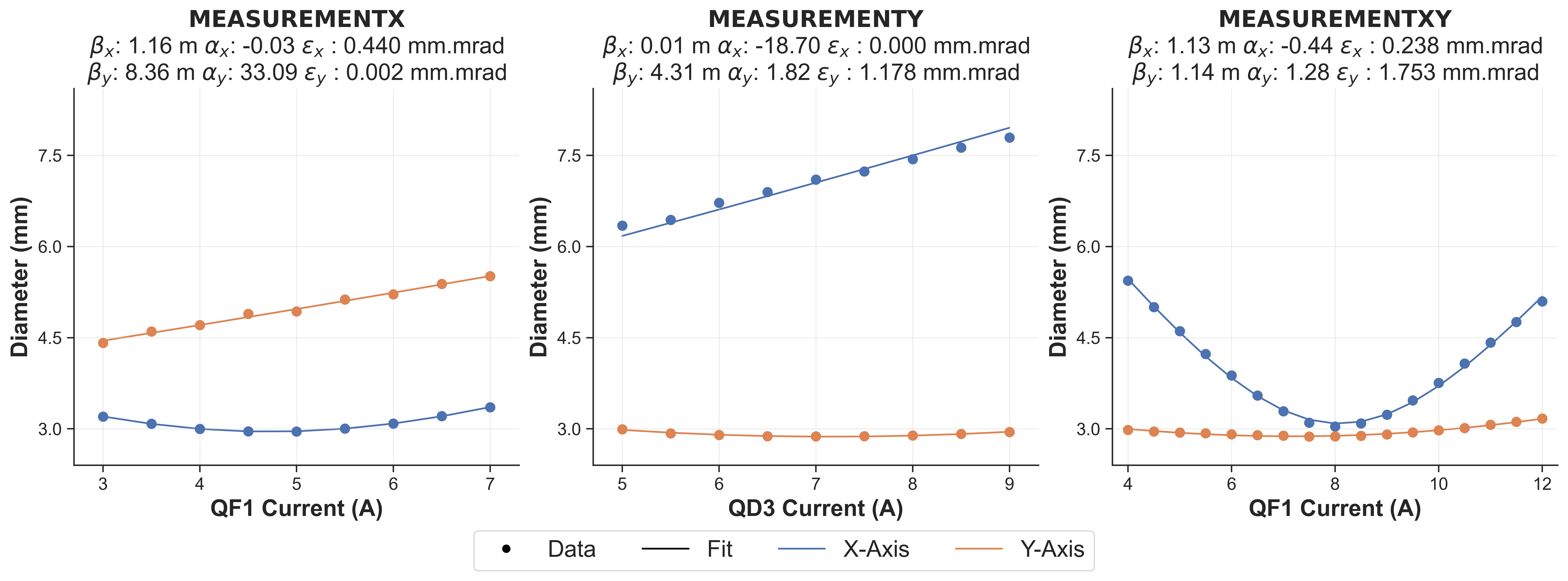}
    \caption{PEER quadrupole scans comparing the measured and fitted beam-diameter curves. An emittance value of 0.000 indicates an emittance that is small but non-zero, rounded to three decimal places.}
    \label{fig:peer_result}
\end{figure}

For the PEER measurements, the method successfully estimated the CS parameters in the $x$-axis for \texttt{MEASUREMENT\_X}, the $y$-axis for \texttt{MEASUREMENT\_Y}, and both transverse axes for \texttt{MEASUREMENT\_XY}. The converged emittance estimates were $0.426$~mm.mrad in the $x$-axis for \texttt{MEASUREMENT\_X}, $1.138$~mm.mrad in the $y$-axis for \texttt{MEASUREMENT\_Y}, and $0.231$~mm.mrad and $1.694$~mm.mrad in the $x$- and $y$-axes, respectively, for \texttt{MEASUREMENT\_XY}. The measured and fitted curves are shown in \autoref{fig:peer_result}, and the estimated CS parameters are summarised in \autoref{tab:peer_result}.

\begin{table}[!htb]
    \centering
    \begin{tabular}{lccc}
        \hline
        \textbf{Parameter} & \texttt{MEASUREMENT\_X} & \texttt{MEASUREMENT\_Y} & \texttt{MEASUREMENT\_XY} \\
        \hline
        $\boldsymbol{\beta_x}$ [m]
            & $1.16$
            & $0.01$
            & $1.13$ \\
        $\boldsymbol{\alpha_x}$
            & $-0.03$
            & $-18.70$
            & $-0.44$\\
        $\boldsymbol{\varepsilon_x}$ [mm.mrad]
            & $0.440$
            & $0.000$
            & $0.238$ \\
        \hline
        $\boldsymbol{\beta_y}$ [m]
            & $8.36$
            & $4.31$
            & $1.14$ \\
        $\boldsymbol{\alpha_y}$
            & $33.09$
            & $1.82$
            & $1.28$ \\
        $\boldsymbol{\varepsilon_y}$ [mm.mrad]
            & $0.002$
            & $1.178$
            & $1.753$ \\
        \hline
        \textbf{RMSE Fit, $x$} [$\mu$m]
            & $2.19$
            & $100.42$
            & $38.01$ \\
        \textbf{RMSE Fit, $y$} [$\mu$m]
            & $29.15$
            & $6.69$
            & $9.96$ \\
        \hline
        \hline
    \end{tabular}
    \caption{CS parameters estimated from the PEER measurements. RMSE values are calculated between the measured and fitted beam diameter curves. \texttt{MEASUREMENT\_X} gives invalid estimates for the $y$-axis, while \texttt{MEASUREMENT\_Y} gives invalid estimates for the $x$-axis. In contrast, \texttt{MEASUREMENT\_XY} provides estimates for both transverse axes.}
    \label{tab:peer_result}
\end{table}

The $x$-axis estimates from \texttt{MEASUREMENT\_X} and \texttt{MEASUREMENT\_XY} yielded similar $\beta$ values but differed in $\alpha$ and emittance. For the $y$-axis, the estimates from \texttt{MEASUREMENT\_Y} and \texttt{MEASUREMENT\_XY} yielded similar $\alpha$ values but differed in $\beta$ and emittance. The measured beam diameter curves were generally less symmetric than those obtained from the simulated quadrupole scans.

\begin{figure}[ht]
    \centering
    \includegraphics[width=0.6\linewidth]{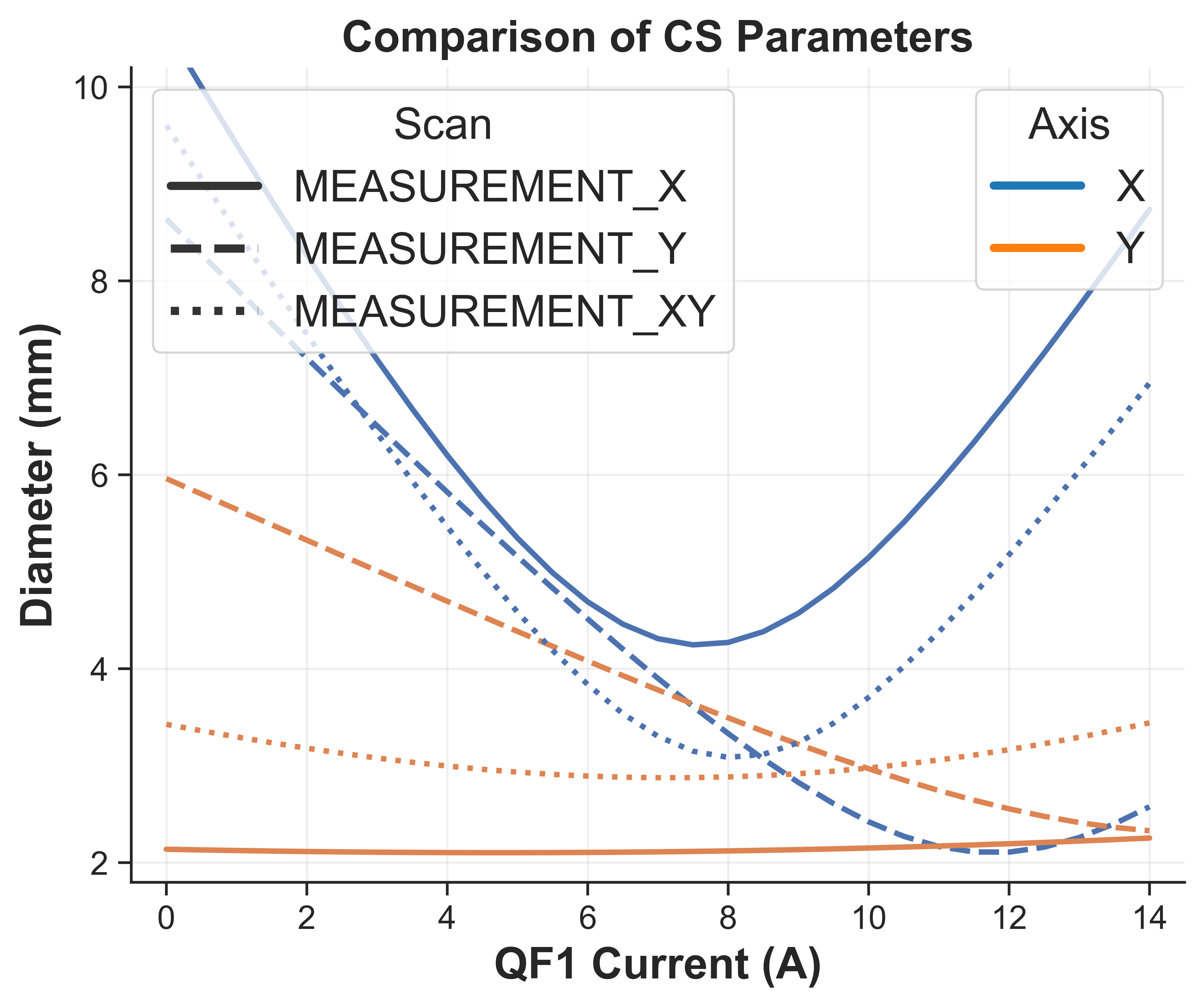}
    \caption{Beam diameter curves obtained using the CS parameters estimated from the three PEER measurements. QF1 is scanned from 0~A to 14~A, QD2 is set to 9~A, and QD3 is switched off.}
    \label{fig:comparison}
\end{figure}

The consistency between the estimates from different scans was further assessed by comparing the beam diameter curves obtained by propagating the CS parameter estimates from each scan through the extended \texttt{MEASUREMENT\_XY} configuration, with QF1 scanned from 0~A to 14~A. The results are shown in \autoref{fig:comparison}. The $x$-axis curves based on \texttt{MEASUREMENT\_X} and \texttt{MEASUREMENT\_XY} exhibit minima near 8~A, but differ in their estimated beam sizes across the scan range. Meanwhile, the $y$-axis curves does not show any similarity between the scans.

\section{Discussion}
\label{discussion}
The scattering model produced estimates that were generally consistent with the MC results across the simulated beam parameters and materials. The fitted correction factors reduced the estimation errors and remained close to unity, suggesting that the original characteristic scattering angle formula provides a reasonable approximation over the investigated parameter space. An additional advantage of the CMF model incorporating MCS is that a correction factor determined for one material may also be applicable to another material with similar scattering properties. For example, water and polymethyl methacrylate (PMMA) have similar scattering properties, suggesting that the correction factor fitted for water may also be suitable for PMMA, although this requires further validation. The correction factors had little effect on the mean kinetic energy because they modify only the characteristic scattering angle, whereas the kinetic energy loss is determined solely by the stopping power.

For all six simulated quadrupole scan configurations, a beam size minimum was observed in both transverse axes, allowing the optimisation to converge to estimates of the initial CS parameters. Incorporating MCS into the CMF model reduced the estimation errors by several orders of magnitude. However, considerable errors remained in some cases, particularly for scans with steeper gradients (more pronounced curvature in the beam size curves). These results indicate that a pronounced variation in beam size does not lead to better estimates. For example, the current range in \texttt{Q2SCANX} was designed to produce substantial changes in the beam size along the $x$-axis, yet the estimated $x$-axis parameters were less accurate than those for the $y$-axis. A similar pattern was observed for \texttt{Q2SCANY}, which yielded more accurate CS parameter estimates for the $x$-axis than for the $y$-axis.

Although the RMSE between the simulated data and fitted curves was at most $6~\mu\mathrm{m}$, the estimated emittances differed from their true values by up to approximately 40\%. This demonstrates that a close fit to the simulated beam sizes does not necessarily guarantee accurate recovery of the initial CS parameters. Furthermore, the fitted curves obtained using the CMF model incorporating MCS generally agreed more closely with the simulated data than the curves generated using the true initial CS parameters. This suggests that discrepancies may persist even when the true parameters are used, potentially because of limitations in the processing of the phase space data. For example, when the beam becomes very small, the histogram bins may be too coarse to support a reliable Gaussian fit, potentially biasing the calculated beam size.

For the PEER measurements, the optimisation generally converged to CS parameter estimates for at least one transverse axis in each scan. These results highlight the importance of a discernible beam size minimum for reliably estimating the corresponding CS parameters. Although the fitted beam sizes may agree with the measurements, monotonic beam size curves provided insufficient constraints on the parameters. A single scan may therefore provide reliable CS parameter estimates for only one transverse axis, making separate scans targeting each axis necessary to characterise both axes. However, a single scan can constrain both axes simultaneously if the scan range captures a beam size minimum in each axis \citep{prat_quadscan_2014}, although achieving this condition can be difficult in practice.

The $x$-axis emittance estimates were consistently smaller than the corresponding $y$-axis estimates when comparing \texttt{MEASUREMENT\_X} and \texttt{MEASUREMENT\_Y} with \texttt{MEASUREMENT\_XY}. Such asymmetry would be unexpected for an ideal linac, where the emittances in the two transverse axes would generally be similar. Possible explanations include asymmetries in the linac or systematic errors in calibrating images of the tilted screen, either of which could produce apparent differences between the transverse axes. However, the substantial differences between the CS parameter estimates from \texttt{MEASUREMENT\_X} and \texttt{MEASUREMENT\_Y} and those from \texttt{MEASUREMENT\_XY} cannot be attributed confidently to a single cause.

Furthermore, the beam profile images indicate beam rotation as the quadrupole current varies, suggesting transverse coupling that may arise from upstream solenoids or a tilted quadrupole. Although a two-dimensional Gaussian fit to the beam profile can account for this rotation through the $x$--$y$ covariance term, the coupling may still have influenced the measured beam sizes and the resulting CS parameter estimates. Given the apparent significance of this effect, future work could incorporate transverse coupling into the optimisation to better constrain the solution and estimate the full initial $4\times4$ covariance matrix, rather than optimising each transverse axis independently.

Although the estimated CS parameters reproduce the measured beam size curves, they remain inconsistent across scans. As shown by the comparison of beam diameter curves obtained using the CS parameters from the three PEER measurements in \autoref{fig:comparison}, each parameter set produces a different estimated beam diameter curve for the same extended \texttt{MEASUREMENT\_XY} quadrupole configuration. The emittance estimates also varied considerably between scans, despite the expectation that the linac output would remain relatively stable over the measurement period. Variations in operating conditions, such as RF power and bunker temperature, could affect cavity resonance and beam acceleration, thereby influencing the CS parameters through dispersion or energy spread. However, because the scans were conducted within approximately one hour of one another, these factors are less likely to explain the magnitude of the observed variation.

These findings indicate that beam characterisation across the usable linac configurations is necessary before operation. For potential clinical applications, this process could be integrated into routine linac quality assurance to verify beam stability and ensure reliable and accurate beam delivery. For example, beam characterisation could be performed alongside the daily quality assurance checks routinely conducted on clinical medical linacs at the start of each day. Further work is required to establish the repeatability of the proposed method and identify potential sources of variation in the beam parameters.

In practice, the true CS parameters may always remain hidden because of the various limitations discussed above. The primary objective is therefore to estimate the beam size accurately under relevant quadrupole configurations. A set of CS parameters that differs slightly from the true values may still be adequate, provided that it reproduces the beam size accurately. However, substantial discrepancies between the estimated and true CS parameters may lead to inaccurate beam size estimations outside the scanned current range, for example, when a different quadrupole is used for scanning. This limitation could be addressed by performing more comprehensive scans using all available quadrupoles, albeit at the cost of additional measurement time.

Nevertheless, the estimated emittances are of the same order of magnitude as the linac design value of approximately $0.8~\mathrm{mm\,mrad}$, corresponding to a normalised emittance of $50\pi~\mathrm{mm\,mrad}$, as reported by \citet{dowd_linac_2011}. Previous measurements using the in-vacuum screen in the Linac-to-Booster (LTB) transfer line also reported emittances of the same order of magnitude, although dispersion introduced by the dipole magnet limits these measurements to the $y$-axis only \citep{meier_linac_2008}. To date, no in-vacuum screen has been installed along the PEER beamline, preventing independent and direct validation of the method presented here.

The primary role of the injector linac is to supply electrons to the synchrotron booster ring rather than to maintain specific CS parameters for radiotherapy applications. Provided that the beam remains sufficiently small and contained during transport through the linac and transfer line, it can still be accelerated to 3~GeV for injection into the storage ring. Consequently, detailed measurements of the transverse beam phase space are not routinely required for the linacs primary operational purpose.

\section{Conclusion}
\label{conclusion}

A CMF model incorporating MCS was developed to estimate VHEE beam parameters from in-air screen measurements downstream of a scattering window. Material-specific correction factors for the characteristic scattering angle were fitted to MC simulations, reducing the estimation errors compared with those obtained using the uncorrected model. In simulated quadrupole scans, incorporating MCS into the CMF model reduced the emittance estimation errors by up to two orders of magnitude. The simulated scans also demonstrated that the optimisation could recover the initial CS parameters more accurately when the beam-size curves varied smoothly and exhibited clearly resolved minima. However, accurate reproduction of the beam-size curves did not necessarily guarantee accurate recovery of the underlying CS parameters.

In-air screen measurements at PEER demonstrated the feasibility of characterising VHEE beams without an in-vacuum screen. The estimated beam sizes generally agreed with the measurements, and the inferred emittances were of the same order of magnitude as previously reported values. Nevertheless, substantial scan-to-scan variation in the estimated CS parameters persisted, and the absence of clearly resolved beam size minima could lead to unreliable parameter estimates. Although the estimated parameters may still provide useful beam size estimations, their consistency and transferability across scan configurations require further investigation.

Potential sources of uncertainty include inaccuracies in the beamline model and quadrupole characteristics, limitations in the spatial resolution of the screen and camera system, and the effects of transverse coupling, dispersion and energy spread, which are not explicitly accounted for in the CMF with MCS model. Further work is therefore required to establish the repeatability and accuracy of the method, characterise the spatial resolution of the in-air imaging system, and investigate these sources of uncertainty.

\section{Acknowledgements}
\label{acknowledgements}

This work is supported by Melbourne Research Scholarship. The authors gratefully acknowledge the assistance of the Australian Synchrotron operators during beamtime.

\clearpage
\appendix
\section{Conventional quadrupole scan}
\label{app:conventional_quadscan}

In the conventional quadrupole scan method, a single focusing quadrupole is followed by a drift space. The initial beam covariance matrix can be reconstructed by fitting a second-order polynomial to the measured downstream beam size as a function of quadrupole focusing strength $k$, which is related to the magnetic field gradient $G$ and the beam rigidity $B\rho$ by
\begin{equation}
    k^2 = \frac{G}{B\rho} = \frac{c}{p}QI,
\end{equation}
where $G$ is the quadrupole magnetic field gradient in $\mathrm{T~m^{-1}}$ and $B\rho$ is the magnetic rigidity in $\mathrm{T~m}$. Alternatively, the focusing strength can be expressed in terms of the quadrupole excitation constant $Q$ in $\mathrm{T~m^{-1}~A^{-1}}$, the quadrupole current $I$ in $\mathrm{A}$, the beam momentum $p$ in $\mathrm{eV}/c$, and the speed of light $c$ in $\mathrm{m~s^{-1}}$.

The polynomial follows from the system transfer matrix, $M(k)$, obtained by modelling the quadrupole as a thin lens with length $L$, followed by a vacuum drift of length $D$: 
\begin{equation}
    M(k) =
    \begin{pmatrix}
        1 & D \\
        0 & 1
    \end{pmatrix}
    \begin{pmatrix}
        1 & 0 \\
        -kL & 1
    \end{pmatrix}
    =
    \begin{pmatrix}
        1-kDL & D \\
        -kL & 1
    \end{pmatrix}.
\end{equation}

Transforming initial covariance matrix $\boldsymbol{C}$ using this transfer matrix and considering only the downstream position variance $C_{11}'$ gives
\begin{equation}
    \begin{aligned}
        C_{11}'(k)
        &= (1 - kDL)^2)C_{11} + 2D(1 - kDL)C_{12} + D^2C_{22} \\
        &= D^2L^2C_{11}k^2 - 2DL(C_{11} + DC_{12})k + C_{11} + 2DC_{12} + D^2C_{22} \\
        &= a_2 k^2 + a_1 k + a_0,
    \end{aligned}
\end{equation}
where $a_0$, $a_1$, and $a_2$ are the fitted polynomial coefficients. The initial covariance matrix elements are recovered from these coefficients as
\begin{equation}
    \begin{aligned}
        C_{11} &= \frac{a_2}{D^2L^2}, \\
        C_{12} &= -\frac{a_1}{2D^2L} + \frac{C_{11}}{D}, \\
        C_{22} &= \frac{a_0 - C_{11}}{D^2} - \frac{2C_{12}}{D}.
    \end{aligned}
\end{equation}
The CS parameters can then be calculated from these elements.

\section{Derivation of the \texorpdfstring{$\boldsymbol{\Delta}$}{Delta} matrix}
\label{app:derive_delta}

Following \citet{negra_mcs_1971}, the scattering covariance matrix $\boldsymbol{\Delta}$ is obtained by integrating the contributions of successive small-angle scattering kicks across the scatterer thickness:
\begin{equation}
    \boldsymbol{\Delta}
    =
    \int_0^L T(\ell)
    \begin{pmatrix}
        (L-\ell)^2 & L-\ell \\
        L-\ell & 1
    \end{pmatrix}
    \mathrm{d}\ell,
\end{equation}
where $\ell$ is the depth within the scatterer and $T(\ell)$ is the local angular scattering variance per unit length. The local scattering variance is assumed to scales inversely with the square of the particle momentum, so the local and initial scattering variances are then related by
\begin{equation}
    T(\ell)
    = T(0)\left(\frac{p(0)}{p(\ell)}\right)^2.
\end{equation}

The initial scattering variance per unit length is related to the characteristic scattering angle $\theta_0$ as follows:
\begin{equation}
    T(0) = \frac{\theta_0^2}{L},
\end{equation}
where $\theta_0$ is given by the MCS expression from \citet{lynch_mcs_1991}:
\begin{equation}
    \theta_0 =
    \lambda
    \frac{13.6z}{pc\beta_{\mathrm{rel}}}
    \sqrt{\frac{L}{X_0}}
    \left[
        1+0.088\log_{10}
        \left(
            \frac{Lz^2}{X_0\beta_{\mathrm{rel}}^2}
        \right)
    \right].
\end{equation}
Here, $p$ is the beam momentum in $\mathrm{MeV}/c$, $c$ is the speed of light in vacuum, $\beta_{\mathrm{rel}}$ is the ratio of particle speed to the speed of light, $z$ is the particle charge number, $L$ is the scatterer thickness, and $X_0$ is the radiation length of the material, expressed in the same length units as $L$. The material-specific factor $\lambda$ is introduced here to account for small deviations between the scattering model and the MC simulations. For the electron ($z=1$) energies considered here, $\beta_{\mathrm{rel}}\approx1$. Radiation lengths were obtained from the Particle Data Group (PDG) database \citep{pdg_2026}.

Momentum loss is approximated as a linear decrease with depth:
\begin{equation}
    p(\ell) = p(0) - a\ell,
\end{equation}
where $a$ is the momentum loss per unit length. The fractional momentum loss, $\delta$, is defined as
\begin{equation}
    \delta
    \equiv
    \frac{p(0)-p(L)}{p(0)}
    =\frac{p_0-p_f}{p_0},
\end{equation}
where $p_0$ and $p_f$ are the initial and final momenta, respectively. Therefore
\begin{equation}
    a=\frac{\delta p_0}{L},
\end{equation}
hence
\begin{equation}
    p(\ell)=p_0\left(1-\frac{\delta\ell}{L}\right).
\end{equation}
Substituting this momentum expression into the scattering covariance integral gives
\begin{equation}
    \boldsymbol{\Delta}
    =
    \frac{\theta_0^2}{L}
    \int_0^L
    \frac{1}{(1-\delta\ell/L)^2}
    \begin{pmatrix}
        (L-\ell)^2 & L-\ell \\
        L-\ell & 1
    \end{pmatrix}
    \mathrm{d}\ell.
\end{equation}
The individual matrix elements are consequently
\begin{equation}
    \Delta_{11}
    =
    \frac{\theta_0^2}{L}
    \int_0^L
    \frac{(L-\ell)^2}
    {(1-\delta\ell/L)^2}
    \mathrm{d}\ell
    =
    \theta_0^2L^2
    \left[
        \frac{
            \delta(2-\delta)
            +2(1-\delta)\ln(1-\delta)
        }{\delta^3}
    \right],
\end{equation}

\begin{equation}
    \Delta_{12}
    =
    \frac{\theta_0^2}{L}
    \int_0^L
    \frac{L-\ell}
    {(1-\delta\ell/L)^2}
    \mathrm{d}\ell
    =
    \theta_0^2L
    \left[
        -\frac{\delta+\ln(1-\delta)}{\delta^2}
    \right],
\end{equation}

\begin{equation}
    \Delta_{22}
    =
    \frac{\theta_0^2}{L}
    \int_0^L
    \frac{1}
    {(1-\delta\ell/L)^2}
    \mathrm{d}\ell
    =
    \theta_0^2\left(\frac{1}{1-\delta}\right).
\end{equation}

To avoid numerical instability as $\delta\to0$, the expressions are expanded as power series in $\delta$. Retaining terms up to the second-order gives

\begin{equation}
    \Delta_{11}
    =
    \theta_0^2L^2
    \left(
        \frac{1}{3}
        +\frac{\delta}{6}
        +\frac{\delta^2}{10}
    \right),
\end{equation}

\begin{equation}
    \Delta_{12}
    =
    \theta_0^2L
    \left(
        \frac{1}{2}
        +\frac{\delta}{3}
        +\frac{\delta^2}{4}
    \right),
\end{equation}

\begin{equation}
    \Delta_{22}
    =
    \theta_0^2
    \left(
        1+\delta+\delta^2
    \right).
\end{equation}
These approximations remain finite as the momentum loss approaches zero while retaining the second-order dependence on $\delta$.

The fractional momentum loss $\delta$ is calculated from the continuous slowing-down approximation (CSDA) range for the material and initial beam kinetic energy $E_0$. The initial CSDA range is subtracted by the scatterer thickness $L$, and the corresponding final kinetic energy $E_f$ is obtained using the inverse range--energy relationship:
\begin{equation}
    E_f=R^{-1}\left(R(E_0)-L\right)
\end{equation}
where $R$ and $R^{-1}$ are the interpolated function relating kinetic energy to CSDA range and CSDA range back to kinetic energy, respectively. The range--energy relationship is obtained from the stopping power tables provided by NIST \citep{berger_nist_1993}. The final momentum is then calculated from $E_f$ and used to determine $\delta$.


\clearpage
\bibliographystyle{elsarticle-num-names} 
\bibliography{refs}




% %% For authoryear reference style

% \bibitem[Lamport(1994)]{lamport94}
%   \textit{\LaTeX: a document preparation system},
%   Addison Wesley, Massachusetts,

% \end{thebibliography}
\end{document}